\documentclass[11pt]{article}

\usepackage[final]{latex/acl}
\usepackage{times}
\usepackage{latexsym}
\usepackage{amsmath} 
\usepackage{multirow} 
\usepackage{multicol} 
\usepackage{graphicx}
\usepackage{xspace}
\usepackage{pdflscape, amssymb, url, hyperref, enumitem, graphicx, mathtools}

\usepackage{booktabs}

\usepackage{algorithm}
\usepackage[noend]{algpseudocode} 
\usepackage{xcolor}
\usepackage{setspace} 
\usepackage{makecell}

\definecolor{lightblue}{RGB}{203, 220, 235}
\definecolor{lightgreen}{RGB}{219,234,210}
\definecolor{lightred}{RGB}{255,130,130}
\definecolor{lightyellow}{RGB}{255, 254, 200}

\definecolor{commentgray}{gray}{0.4}

\newcommand{\argmax}{\operatorname{argmax}}
\newcommand{\KwTo}{\leftarrow}
\usepackage[T1]{fontenc}

\usepackage[utf8]{inputenc}

\usepackage{microtype}

\usepackage{inconsolata}

\title{IKFST: IOO and KOO Algorithms for Accelerated and Precise WFST-based End-to-End Automatic Speech Recognition}

\author{Zhuoran Zhuang\thanks{Equal contribution.}$^{\clubsuit}$, 
Ye Chen$^{*\clubsuit}$,
Chao Luo$^{*\clubsuit}$,
Tian-Hao Zhang$^{*\spadesuit}$,\\
\textbf{
Xuewei Zhang$^{\clubsuit}$,
Jian Ma$^{\clubsuit}$,
Jiatong Shi$^{\heartsuit}$,
Wei Zhang\thanks{Corresponding author.}$^{\clubsuit}$
}\\
$^{\clubsuit}$ Fliggy Alibaba\\
$^{\spadesuit}$ University of Science and Technology Beijing\\
$^{\heartsuit}$Carnegie Mellon University\\
\texttt{zhichen.zw@alibaba-inc.com, tianhaozhang@xs.ustb.edu.cn}
}

\begin{document}
\maketitle
\begin{abstract}

End-to-end automatic speech recognition has become the dominant paradigm in both academia and industry. To enhance recognition performance, the Weighted Finite-State Transducer (WFST) is widely adopted to integrate acoustic and language models through static graph composition, providing robust decoding and effective error correction. However, WFST decoding relies on a frame-by-frame autoregressive search over CTC posterior probabilities, which severely limits inference efficiency.
Motivated by establishing a more principled compatibility between WFST decoding and CTC modeling, we systematically study the two fundamental components of CTC outputs, namely blank and non-blank frames, and identify a key insight: blank frames primarily encode positional information, while non-blank frames carry semantic content.
Building on this observation, we introduce Keep-Only-One and Insert-Only-One, two decoding algorithms that explicitly exploit the structural roles of blank and non-blank frames to achieve significantly faster WFST-based inference without compromising recognition accuracy.
Experiments on large-scale in-house, AISHELL-1, and LibriSpeech datasets demonstrate state-of-the-art recognition accuracy with substantially reduced decoding latency, enabling truly efficient and high-performance WFST decoding in modern speech recognition systems.
\vspace{-2mm}
\end{abstract}

\section{Introduction}
\label{Intro}
\vspace{-2mm}

Recent advances in deep neural networks have substantially improved automatic speech recognition (ASR) \cite{ref1, ref2, ref3, ref4, ref5}. Traditional ASR systems rely on multi-stage pipelines with expert-designed components \cite{ref6}, while end-to-end (E2E) models simplify this process by directly mapping speech to text \cite{ref7}. E2E ASR methods mainly include CTC \cite{ref8, ref9, ref10}, RNN-T \cite{ref11, ref12}, and attention-based encoder–decoder models \cite{ref13, ref14, ref15}, as well as recent hybrid formulations such as CTC/RNN-T and CTC/AED \cite{ref16, ref17, ref18}. In these systems, external language models are commonly integrated during inference to enhance linguistic consistency and domain adaptation, typically through N-best re-ranking or non-autoregressive re-scoring methods \cite{ref19, ref20, ref21, ref22}.

Among various decoding frameworks, the Weighted Finite-State Transducer (WFST) has long served as a foundational component in large-scale ASR systems, owing to its strong theoretical grounding, flexible topology composition, and robust decoding capability \cite{ref23, ref24, ref25}. Within this framework, CTC models provide frame-level posterior probabilities, which are then decoded by a modified WFST through Viterbi beam search \cite{ref26, ref27, ref28}. Despite its effectiveness and widespread adoption, directly performing WFST decoding over the full posterior sequence remains computationally prohibitive, posing a critical challenge to efficient and scalable deployment.
To alleviate the aforementioned issues, Chen et al \cite{ref29} proposed the Lattice-based Phone Synchronous Decoding (LSD) algorithm. SpeechLLaMA \cite{ref30} proposed an averaging strategy to preserve this latent information. PolyVoice \cite{ref31} systematically discards all blank frames. Building on similar insights, the Spike Window Decoding (SWD) algorithm further refined this concept by selectively incorporating a limited neighborhood of blank frames surrounding high-probability non-blank spikes \cite{ref32}.

From the collective insights of prior research, a unifying observation arises: even after pruning a significant number of blank frames, the model can deliver competitive or even superior recognition results in the WFST decoding. This observation motivates a deeper investigation into the intrinsic positional and semantic characteristics of blank and non-blank frames, aiming to uncover the minimal yet sufficient frame representation for optimal decoding.
Building on this perspective, we introduce two complementary algorithms: Insert-Only-One (IOO) and Keep-Only-One (KOO). The IOO algorithm operates by first discarding all probabilistic, model-learned blank frames and then strategically inserting a deterministic, user-defined blank frame between adjacent non-blank frames thereby preserving crucial transitional cues while simultaneously eliminating superfluous temporal redundancy. In parallel, the KOO algorithm addresses redundancy within the non-blank domain by selectively retaining only one representative spike per activation cluster, effectively compressing the posterior sequence without degrading its semantic fidelity. Notably, the IOO algorithm is broadly applicable, enhancing the performance of both CTC-FST and AED-FST decoding. 

We conduct an extensive evaluation of the proposed KOO and IOO algorithms across diverse datasets to thoroughly assess their effectiveness and generalization. Experiments are performed on the widely adopted AISHELL-1 Mandarin dataset \cite{ref33}, the English Librispeech dataset \cite{ref34}, as well as on a large-scale 65,000-hours In-House dataset. We first construct a CTC/AED hybrid acoustic model combined with a GPU-accelerated WFST decoding framework \cite{ref35}, forming a robust foundation that achieves state-of-the-art (SOTA) recognition accuracy.
Building on this, the proposed methods demonstrate significant gains in both inference speed and recognition performance. These results validate the approach's capability to deliver efficient, high-accuracy decoding across diverse linguistic contexts and data scales.

\section{Related Work}
\label{Related work}

\subsection{CTC-based ASR Model}
CTC is a widely adopted objective for end-to-end ASR due to its ability to train models without requiring frame-level alignments. 
For a given an acoustic feature sequence $X$, we define $Y$ as the corresponding label sequence, which has a length of $L$. the encoder produces a sequence of hidden representations:
\begin{align}
    H_{encoder} = \mathrm{Encoder}(X).
\end{align}
CTC introduces an intermediate alignment sequence by allowing blank symbols and repeated tokens, enabling flexible many-to-one mappings between acoustic frames and output labels.  
Let $\mathcal{B}(\cdot)$ denote the mapping that removes blanks and repeated symbols; the CTC objective is:
\begin{align}
    \mathcal{L}_{CTC}
    = - \log \sum_{Z \in \mathcal{B}^{-1}(Y)}
        p(Z \mid H_{encoder}),
\end{align}
where the probability of an alignment path $Z$ is modeled under conditional independence assumptions across time steps.


\subsection{Hybrid CTC/AED Algorithm}
CTC is frequently combined with an attention-based encoder–decoder (AED). 
The AED decoder conditions on both encoder representations and previously generated tokens:
\begin{align}
    H_{decoder}
        = \mathrm{Decoder}(H_{encoder}, Y),
\end{align}
and is trained using cross-entropy:
\begin{align}
    \mathcal{L}_{AED}
        = \mathrm{CrossEntropy}(H_{decoder}, Y).
\end{align}
The hybrid objective interpolates the two losses:
\begin{align}
    \mathcal{L}
    = \alpha\, \mathcal{L}_{CTC}
    + (1 - \alpha)\, \mathcal{L}_{AED},
\end{align}
where $\alpha$ is a hyper-parameter used to adjust the weight ratio between the encoder
and decoder, with its value ranging from $[0, 1]$, which is routinely configured to 0.1 for the rest of the study.

\begin{figure*}[!th]
  \centering
  \includegraphics[width=0.95\linewidth]{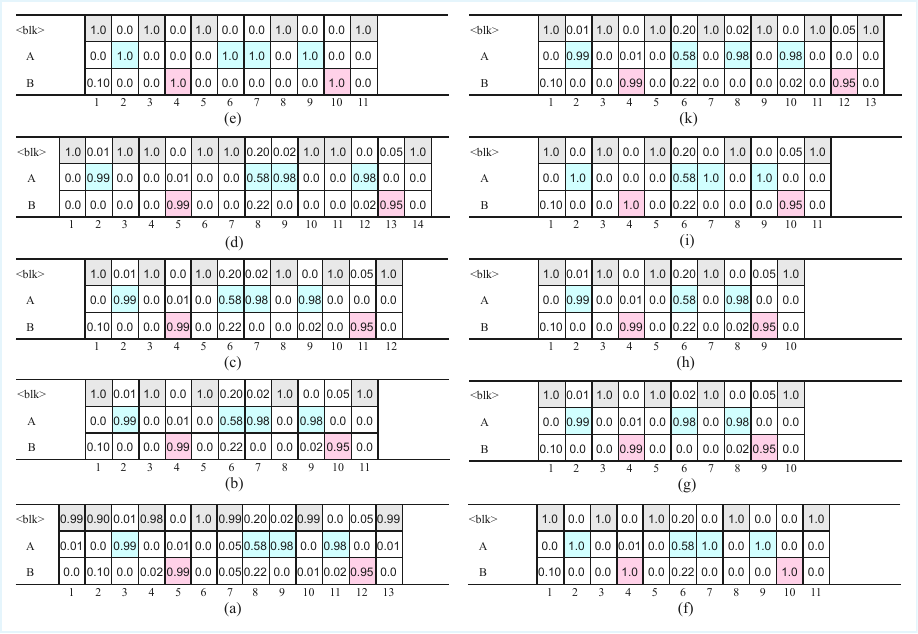}
  \caption{(a) represents a search performed using dense frames; (b)-(d) and (e)-(f) illustrate the use of the IOO algorithm on blank frames and non-blank frames, respectively; (g) preserves the frame with the highest posterior probability, whereas (h) retains the frame with the lowest posterior probability; (i) denotes the application of the KOO algorithm based on probability; (k) indicates the integration of AED-FST with the IOO algorithm.}
\label{figure1}
\end{figure*}

\subsection{WFST based decoding algorithm}
We integrate a WFST-based decoding module on the encoder side and construct a static decoding graph following the standard TLG composition framework. 
The overall graph is obtained through the sequential composition of the token, lexicon, and grammar transducers:
\begin{align}
    \label{eq2}
    TLG = T \circ \mathrm{min}\!\left(\mathrm{det}\!\left(L \circ G\right)\right),
\end{align}
where $L$ and $G$ denote the lexicon FST and the grammar FST, respectively.  
The operators $\mathrm{det}$, $\mathrm{min}$, and $\circ$ correspond to determinization, minimization, and WFST composition.  
It is worth noting that we adopt the GPU-accelerated WFST decoder and the compact token transducer $\mathrm{T}$ proposed in \cite{ref35}, which substantially reduces the state-space complexity of graph construction—from second-order exponential to linear, while preserving competitive decoding accuracy.

\section{Methodology}
\label{method}
In this study, we investigate the integration of CTC posterior probabilities with FST to achieve optimal trade-offs between inference speed and recognition accuracy. We systematically analyze the distinct characteristics of blank and non-blank outputs within the CTC posterior, and based on this, propose two novel algorithms: Insert-Only-One (IOO) and Keep-Only-One (KOO). By combining these two algorithms, we demonstrate a significant improvement in both processing speed and recognition performance. Notably, using the IOO algorithm is an effective way to mitigate the severe drop in accuracy for AED-FST. 

\subsection{IOO and KOO Algorithms}
In the CTC framework, the encoder produces a sequence of logits $H_{encoder}$, which represent the raw, unnormalized token scores at each timestep. To obtain the corresponding posterior logits $P$, the softmax function is applied over the vocabulary dimension:
\begin{align}
P = \mathrm{SoftMax}(H_{encoder}).
\end{align}
This yields a probability distribution over all possible tokens which including the blank symbol at every frame. As illustrated in Fig. \ref{figure1} (a), the posterior distribution produced by the CTC model exhibits a pronounced spike-like pattern, characterized by a large number of high-probability blank predictions interspersed with token activations (represented in the figure using two example token classes A and B). When the decoding process is performed in a dense manner, that is, by providing the complete CTC posterior sequence directly to the FST, the decoding efficiency degrades substantially. This degradation is caused by the overwhelming number of blank-dominated frames, which greatly expand the effective search space and introduce significant computational overhead during FST traversal.

Therefore, the IOO algorithm is introduced to mitigate this issue. As illustrated in Fig.\ref{figure1} (b), the IOO algorithm performs a controlled insertion of customized blank frames. This strategy preserves the desirable temporal separation provided by blank symbols while simultaneously alleviating the decoding inefficiency caused by densely occurring blank regions in the original CTC posterior sequence. Specifically, given the CTC posterior sequence $P$, the IOO algorithm removes all original blank frames and replaces each position containing one or more consecutive blank predictions with a single customized blank distribution. Each customized blank frame is defined as a one-hot probability vector in which the blank token (indexed as 0) is assigned a posterior probability of 1.0, while all remaining token probabilities are set to 0.0. Formally, the inserted blank frame is given by:
\begin{align}
p_{blk} = [1.0,\, 0.0,\, 0.0,\, \ldots,\, 0.0] \in \mathbb{R}^{|\mathcal{V}|},
\end{align}
where $|\mathcal{V}|$ denotes the size of vocabulary. This replacement not only eliminates the inefficiency associated with densely occurring blank segments but also introduces a consistent and deterministic blank representation that improves the stability of FST decoding. The final IOO-enhanced posterior sequence is denoted as $P_{IOO}$.

Complementary to the proposed IOO algorithm, which optimizes redundant blank frames, we further introduce the KOO algorithm to refine the non-blank components of the CTC posterior sequence. Given the frame-level probability sequence $P_{IOO}$, the KOO algorithm identifies non-blank regions directly along the original temporal axis. Let the predicted token at frame $t$ be denoted by $c_t = \arg\max_{v \in \mathcal{V}} P_{t,v}$. After that, the non-blank blocks $B$ are then constructed by grouping consecutive timesteps that share the same non-blank prediction:
\begin{align}
B_k
=
\left\{
t_k^{\mathrm{start}},
t_k^{\mathrm{start}}+1,
\dots,
t_k^{\mathrm{end}}
\right\}, \\
\text{s.t. } c_t = c_k \neq \text{blank},\;
\forall t \in B_k,
\end{align}
where the index $k$ enumerates the non-blank blocks in temporal order, and $t_k^{\mathrm{start}}$ corresponds to the first timestep belonging to the $k$-th block. Importantly, KOO must operate on the full sequence $(c_1, c_2,...,c_T)$ without removing blank frames, discarding blank frames prematurely could artificially merge two non-adjacent frames and produce incorrect temporal adjacency.

As shown in the Fig.\ref{figure1} (g) and Fig.\ref{figure1} (h), for each block $B_k$, KOO selects a single representative frame according to either the maximum- or minimum-probability strategy:
\begin{align}
f^* =
\begin{cases}
\displaystyle \arg\max_{t \in B_k} P_{t,\, c_k}, & \text{max strategy}, \\[8pt]
\displaystyle \arg\min_{t \in B_k} P_{t,\, c_k}, & \text{min strategy}.
\end{cases}
\end{align}
Collecting the logits associated with the selected indices produces the refined non-blank sequence:
\begin{align}
L_{\mathrm{nb}}
=
\left[
P_{f^*_1},
P_{f^*_2},
\dots,
P_{f^*_k}
\right]
\in \mathbb{R}^{K \times |\mathcal{V}|}.
\end{align}
The resulting sequence $L_{\mathrm{nb}}$ preserves the temporal structure of the original CTC output while eliminating the cumulative errors introduced by repeated non-blank frames within the FST decoding process. By retaining only a single representative frame for each non-blank region, the sequence enables a more accurate and reliable decoding outcome.

When applying KOO to the IOO-processed probability sequence $P_{IOO}$, the final posterior $P_{final}$ combines positional information from IOO and semantic information from KOO, which preserves $K$ non-blank frames. Since the IOO algorithm inserts at most one customized blank frame after each retained non-blank frame, the length of the final sequence satisfies:
\begin{align}
\lvert P_{\mathrm{final}} \rvert \leq 2K + 1.
\end{align}
Because $K$ is significantly smaller than the original number of timesteps $T$ in the CTC posterior sequence, the number of frames that participate in the subsequent FST decoding is drastically reduced. As a result, the total number of FST search steps decreases proportionally, leading to substantially improved decoding efficiency. The pseudo code for the proposed IOO and KOO procedures for CTC-FST is provided in Algorithm \ref{alg:1} (Appendix A) .

\subsection{AED-IOO Integration}
Although AED-based speech recognition models have recently emerged as a dominant paradigm and have achieved state-of-the-art performance on many benchmarks, our experiments reveal an important limitation: directly feeding the raw AED  posterior sequence into an FST-based decoder yields unsatisfactory results. We believe that, despite the absence of an explicit blank concept in AED models, the FST search process still relies on a form of temporal separation between adjacent frames to operate effectively.

Motivated by this observation, we extend the IOO procedure to AED by inserting a customized blank token after every decoder-generated frame. The resulting output sequence is illustrated in Fig.\ref{figure1} (k). In practical terms, this corresponds to augmenting the AED posterior sequence with a deterministic blank distribution—identical in form to the custom blank used on the CTC side. Empirically, this modification consistently improves recognition accuracy, demonstrating that the introduction of blank-induced temporal separation provides a beneficial regularization effect for FST decoding, even in AED frameworks that do not natively employ blank symbols. The pseudo code for the proposed IOO algorithm for AED-FST is provided in Algorithm \ref{alg:2} (Appendix B) .

\subsection{TLG graph optimization}
The conventional approach for constructing CTC-based TLG decoding graphs is summarized in Eq.~\ref{eq2}. However, when applied to large-scale language models, this construction process often results in an exceedingly large TLG graph, which in turn leads to substantial memory consumption, increased storage requirements, and a more costly inference procedure. To address these challenges, the present work incorporates a weight-pushing step into the graph-building pipeline, positioned between the $\mathrm{det}$ and $\mathrm{min}$ operations. By shifting the path weights toward earlier states, this operation effectively enables early pruning of low-probability paths without sacrificing decoding accuracy. As a result, both the effective search space and the runtime efficiency are significantly improved. The resulting static graph construction procedure is formalized as follows:
\begin{align}
    \label{eq5}
   TLG = T\circ\mathrm{min}(\mathrm{pushing}({\mathrm{det}}(L\circ G)))
\end{align}

\section{Experiments}
\label{sec:experiments}
\subsection{Dataset}
We conduct experiments on LibriSpeech, AISHELL-1, and large-scale In-House datasets. LibriSpeech is a 960-hour English corpus, while AISHELL-1 is a 178-hour Mandarin corpus recorded at a sampling rate of 16 kHz. Our in-house dataset is also a Mandarin speech corpus, recorded at 8 kHz. It contains approximately 65,000 hours of labeled training data, and the test set consists of around 6,000 randomly sampled short utterances from both inbound and outbound telephone calls.

\subsection{Experimental settings}

\subsubsection{Acoustic model settings}
\begin{table*}[!t]
    \caption{Experiment results on the In-House dataset. The arrows indicate whether higher or lower values are preferable.}
    \centering
    \small
    \renewcommand{\arraystretch}{1.0}
    \label{tab:in-house}
    \setlength{\tabcolsep}{4.5mm}{
    \begin{tabular}{cc ccc}
    \toprule
    \textbf{ID} & \textbf{Model} & \textbf{Decoding type}  & \textbf{CER (\%)} $\downarrow$ & \textbf{Speed up} $\uparrow$ \\
    \midrule
    \textbf{A1} & CTC Zipformer & CTC Greedy Search & 3.27  & 3.45 $\times$ \\
    \textbf{A2} & AED Zipformer & AED Greedy Search & 3.21 & 0.87 $\times$ \\
    \textbf{B1} & A1 + 5-gram & Dense CTC FST &  3.09 & 1.00 $\times$  \\
    \textbf{B2} & A2 + 5-gram & Dense AED FST & 3.56 & 0.51 $\times$ \\
    \midrule
    \textbf{C} & B1 + LSD \cite{ref29} &  0.99 \textit{blank} threshold & 3.12 & 1.18 $\times$ \\
    \textbf{D} & B1 + Speech-LLaMA \cite{ref30} & Averaging &  3.10 &  1.29 $\times$ \\
    \textbf{E} & B1 + PolyVoice \cite{ref31} & Discarding & 4.71 & 1.99 $\times$ \\
    \textbf{F} & B1 + SWD \cite{ref32} & $\mathrm{\{0, \pm\ 1\}}$ & 3.08 & 1.47 $\times$ \\
    \midrule
    \textbf{G1} & B1 + IOO-B & $\mathrm{\{0, 1\}}$ & 3.06 &1.65 $\times$  \\
    \textbf{G2} & B1 + IOO-B & $\mathrm{\{0, 1, 2\}}$ &3.06 & 1.54 $\times$ \\
    \textbf{G3} & G1 + IOO-NB & $\mathrm{\{*\}}$ & 3.78 &  1.72 $\times$ \\
    \textbf{G4} & G1 + IOO-NB & $\mathrm{Max}\mathrm{\{*\}}$ & 3.75& 1.75$\times$ \\
    \textbf{G5} & G1 + KOO & $\mathrm{Min}\mathrm{\{*\}}$ &3.06 & 1.67 $\times$ \\
    \textbf{G6} & G1 + KOO & $\mathrm{Max}\mathrm{\{*\}}$ &\textbf{3.05} & 1.69 $\times$ \\
    \textbf{H} & B2 + IOO-B & $\mathrm{\{0, 1\}}$ & 3.13 & 0.43 $\times$ \\
    \bottomrule
    \end{tabular}}
\end{table*}

The acoustic model constructed in this work is a multi-task Hybrid CTC/AED structure, in which the encoder and decoder models are based on the Zipformer and transformer models, respectively. On the all datasets, the encoder follows the large size Zipformer in \cite{ref36}, the decoder side is based on the transformer standard model, which has 6 transformer blocks. The final size of the acoustic model is 0.22B parameters, and the key encoder-related configurations are summarized in Table 3 (Appendix C) .

\subsubsection{Language model settings}
For the AISHELL-1 and LibriSpeech datasets, the language models are constructed exclusively from the text in their respective training sets. Specifically, 5-gram language models are trained using the SRILM toolkit\footnote{https://www.sri.com/platform/srilm}. On the In-House dataset, we use about 43 million pieces of Mandarin dialog text data to construct the language model. The operations $\mathrm{composition}$, $\mathrm{det}$, $\mathrm{min}$ and $\mathrm{weight \ pushing}$ introduced in Section \ref{method} are implemented using the Openfst tool \footnote{https://www.openfst.org/twiki/bin/view/FST/WebHome}.

\subsubsection{Training, inference and evaluation}
During the training stage, 80-dimensional filter banks are extracted as speech features, with a frame length of 25 ms and a frame shift of 10 ms. To augment the data, a speech speed perturbation \cite{ref37} is used, using perturbation coefficients of 0.9, 1.0, and 1.1. Furthermore, the SpecAugment \cite{ref38} strategy is also used to enhance the robustness of the model. All models are trained on 16 NVIDIA Tesla H200 GPUs with mixed precision training. For inference, all our computations are consistently performed on a Tesla T4 GPU (16GB) with a 16-core CPU and 32GB of RAM. For the inference stage, recognition performance is evaluated using the standard Character Error Rate (CER), computed by measuring the Levenshtein distance \cite{ref39} between the predicted sequence and the corresponding ground-truth transcription. To further accelerate decoding, a batch size of 5 is employed for all experiments.

\subsection{Experimental results}
\subsubsection{Experiment of the In-House dataset.}

Table \ref{tab:in-house} presents a quantitative evaluation of the proposed IOO and KOO algorithms on the In-House dataset, benchmarked against standard decoding baselines and representative heuristic acceleration methods. Rows A1 and A2 report greedy decoding results for the Zipformer model with CTC and AED outputs, respectively, defining the performance bounds under simple greedy search. Rows B1 and B2 show the corresponding results obtained with 5-gram Dense WFST decoding, serving as the primary accuracy–latency references. In particular, Row B1 (CTC Zipformer + 5-gram) establishes the dense baseline, achieving a CER of 3.09\% and a normalized decoding speed of $1.00\times$, against which all subsequent acceleration results are measured.

Experiments C focus on evaluating the impact of a label synchronization-based algorithm when integrated with the TLG decoding graph, which using a 0.99 blank threshold. The LSD algorithm is applied under the hypothesis that as the frame discard threshold increases, recognition accuracy will approach that of the vanilla dense system, but with a corresponding reduction in inference speed. Experiments D and E involve averaging the blank frames between the neighbouring non-blank frames and discarding all the blank frames, respectively. Although these methods do enhance inference speed, they come with a more substantial trade-off in recognition accuracy. 

\begin{figure*}
    \centering
    \includegraphics[width=1\linewidth]{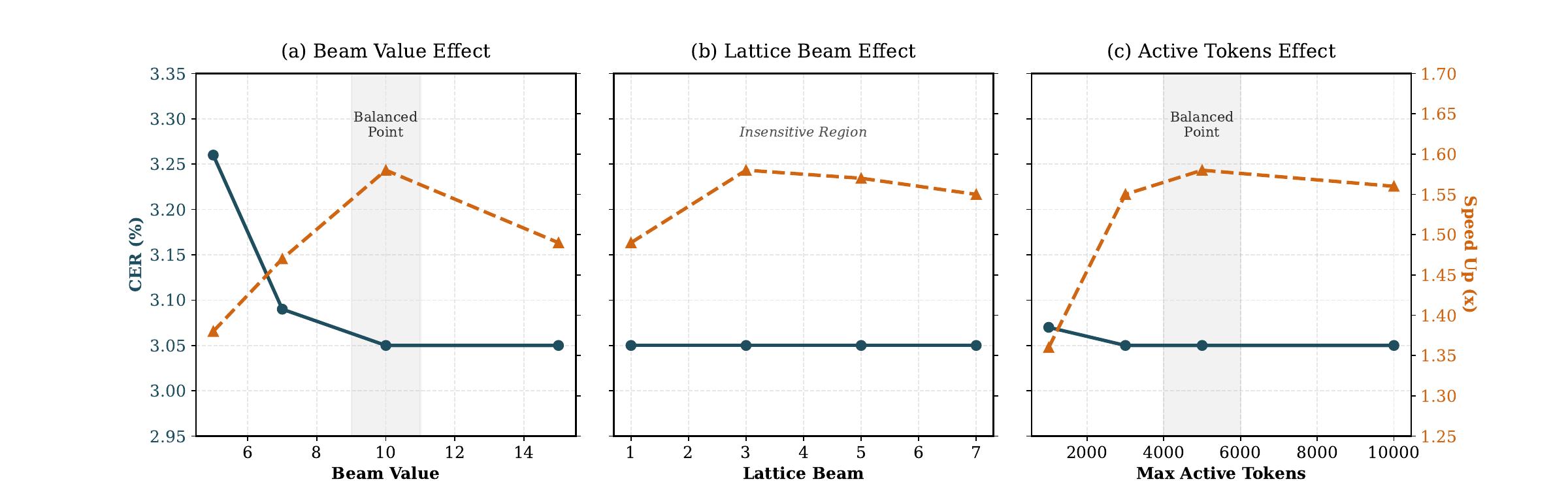}
    \caption{Sensitivity analysis of decoding parameters. The dual-axis plots illustrate the impact of (a) Beam Value, (b) Lattice Beam, and (c) Max Active Tokens on character error rate (CER, solid teal lines) and decoding speed-up (dashed orange lines). Shaded regions indicate the optimal operating points identified in the study, confirming that Lattice Beam is largely insensitive while Beam and Active Tokens require balanced tuning.}
\label{figure2}
\end{figure*}

Starting from the B1 baseline, G1 demonstrates that IOO-B accelerates decoding by 1.65× without any loss in recognition accuracy. Inserting additional blank frames in G2 offers no improvement and slightly reduces speed, while aggressive simplification in G3, replacing all non-blank frames with deterministic one-hot distributions, causes substantial accuracy degradation. A more conservative strategy in G4, retaining the highest-posterior frame within clusters of identical tokens, mitigates accuracy loss while still improving speed.

Experiments G5 and G6 validate the KOO algorithm: retaining only a single non-blank frame per token achieves significant acceleration without degrading accuracy, with the highest-posterior frame yielding the best performance. These results indicate that, for CTC-based WFST decoding, preserving appropriate blank frames and retaining only the most informative non-blank frames maintains recognition accuracy while maximizing computational efficiency.

Experiment H further confirms the generality of IOO on AED outputs, effectively mitigating the accuracy degradation observed with direct WFST decoding. Overall, these findings corroborate the key insight that blank frames encode positional information, whereas non-blank frames carry semantic content.

We further explore the impact of decoding hyperparameters on FST performance. The analyses in Fig. \ref{figure2} indicate that Beam values must balance pruning and search breadth to maintain accuracy and speed, while Lattice Beam can be fixed without affecting performance. Max Active Tokens requires careful tuning to avoid search collapse or excessive memory usage, with around 5000 tokens providing an effective operating point. Importantly, these results confirm that the IOO–KOO framework’s improvements are not contingent on specific parameter choices: a single customized blank per blank region and the highest-probability non-blank frame per spike suffice to preserve accuracy while substantially accelerating decoding.

\begin{table*}[!t]
    \caption{Experiment results on the AISHELL-1 dataset. The results of the open-source dataset are evaluated using CER, while the LibriSpeech dataset is assessed using WER.}
    \centering
    \small
    \renewcommand{\arraystretch}{1.0}
    \label{tabel2}
    \setlength{\tabcolsep}{1.1mm}{
    \begin{tabular}{ccccccc}
    \toprule
     \multirow{2}{*}{\textbf{Model}} & \multicolumn{3}{c}{\textbf{AISHELL-1}} & \multicolumn{3}{c}{\textbf{LibriSpeech}} \\
    & \makecell{Dev (\%)} $\downarrow$ & {Test (\%)} $\downarrow$ & \makecell{Speed up} $\uparrow$ & \makecell{Test-clean (\%)} $\downarrow$ & {Test-other (\%)} $\downarrow$ & \makecell{Speed up} $\uparrow$ \\
    \midrule
    CR-CTC \cite{ref10} & \textbf{3.69} & 3.98 & - & \textbf{1.88} & 3.95  & - \\
    CIF-Transducer \cite{ref18} & 4.1 & 4.3  & - & - & - & - \\
    Zipformer \cite{ref36} & 4.03 & 4.28  & - & 1.96 & 4.08  & -\\
    Branchformer \cite{ref40} & 4.19 & 4.43  & - & 2.4 & 5.3 & - \\
    E-Branchformer \cite{ref41} & 4.2 & 4.5  & - & 2.4 & 4.6 & - \\
    Paraformer \cite{ref42} & 4.7 & 5.1  & - & - & - & - \\
    Conformer \cite{ref43} & 4.5 & 4.9  & - & 2.1 & 4.3  & -\\
    \midrule
    A1 & 3.98 & 4.19  & 3.63 $\times$ & 1.99 & 4.08  & 3.20 $\times$\\
     A2 & 3.97 & 4.15  & 0.90 $\times$  & 2.94 & 4.15  & 0.89 $\times$ \\
      B1 & 3.73 & 3.94  & 1.00 $\times$ & 1.95 & 3.94  & 1.00 $\times$ \\
       B2 & 6.33 & 6.76  & 0.68 $\times$ & 3.77 & 6.98  & 0.55 $\times$ \\
        \midrule
        G1 & 3.71 & 3.93  & 2.37 $\times$ & 1.94 & 3.95  & 2.07 $\times$ \\
         G6 & 3.70 & \textbf{3.92}  & 2.36 $\times$ & 1.94 & \textbf{3.94}  & 2.06 $\times$ \\
          H & 3.77 & 4.03  & 0.55 $\times$ & 2.01 & 4.18  & 0.49 $\times$ \\
    \bottomrule
    \end{tabular}}
\end{table*}

\subsubsection{Experiment of the open-source datasets.}
Table \ref{tabel2} present the experimental results on AISHELL-1 and LibriSpeech datasets. Across both datasets, our IOO–KOO framework consistently improves decoding efficiency while preserving recognition accuracy. On AISHELL-1, applying IOO or KOO to the 5-gram CTC system (B1) yields further CER reductions—from 3.73\%/3.94\% to 3.71\%/3.93\% (G1) and 3.70\%/3.92\% (G6), while achieving over 2.3 times decoding speedup. On LibriSpeech dataset, similar trends are observed. Relative to the dense CTC baseline B1 (1.95\%/3.94\%), both IOO (G1) and KOO (G6) maintain comparable accuracy and deliver more than 2 times speedup. Overall, the results across Mandarin and English corpora confirm that IOO–KOO is robust to language type and corpus scale. The framework consistently delivers substantial decoding acceleration while achieving equal or even improved recognition accuracy, underscoring its practical value for real-world FST-based ASR deployment.



\section{Discussion and Analysis}
\label{discussion}

\textbf{IOO for positional information}
We hypothesize that blank frames in CTC-style formulations serve a dual functional role during FST decoding: beyond absorbing non-semantic acoustic variations, they implicitly encode positional information along the temporal axis, thereby constraining the relative ordering and spacing of non-blank tokens. This hypothesis is supported by several empirical observations. First, in CTC-based models, token sequences such as "A–B–\textless blk\textgreater–\textless blk\textgreater–\textless blk\textgreater", "\textless blk\textgreater–\textless blk\textgreater–\textless blk\textgreater–A–B", and "\textless blk\textgreater\allowbreak–A\allowbreak–\textless blk\textgreater\allowbreak–B\allowbreak–\textless blk\textgreater" are all semantically equivalent to the target sequence "AB", reflecting the alignment-invariant nature of CTC. However, our experimental results show that aggressively discarding blank frames only marginally affects performance, whereas their complete removal leads to a severe degradation in recognition accuracy. Moreover, although AED-based models typically suffer substantial accuracy loss under conventional FST decoding, their performance is significantly improved when integrated with the proposed IOO algorithm, which preserves essential blank-frame structure during decoding.
These findings indicate that blank frames act as an implicit form of positional encoding in FST decoding, anchoring non-blank tokens in time and stabilizing the decoding process. Consequently, inserting at least one blank frame between any two adjacent non-blank frames is crucial for achieving performance comparable to Dense-FST decoding. This insight provides a principled explanation for the necessity of blank frames: they are not merely placeholders for non-semantic acoustic content, but also carry indispensable positional information that governs temporal alignment and decoding robustness.

\textbf{KOO for semantic information}
The KOO algorithm improves FST decoding by selectively discarding semantically redundant non-blank frames, thereby suppressing irrelevant acoustic variations while preserving essential information. This targeted pruning yields higher-quality inputs for subsequent decoding and leads to consistent gains in both efficiency and recognition accuracy. The effectiveness of KOO is supported by three key observations. First, the performance improvements initially observed in Table \ref{tab:in-house} (G5 and G6) are consistently reproduced in Table \ref{tabel2} (G6), demonstrating the robustness of the proposed method. Second, as the pruning threshold increases from 0.8 to 0.99 (Appendix D), recognition accuracy improves monotonically, reaching a relative gain of 3.06\% in G10, while still maintaining over a 1.6× inference speedup. Finally, KOO is theoretically well aligned with the many-to-one mapping property of CTC. During training, adjacent identical non-blank frames are naturally merged within the CTC lattice, and during decoding, sequences such as "A–A–\textless blk\textgreater–\textless blk\textgreater", "\textless blk\textgreater–\textless blk\textgreater–A–A", and "\textless blk\textgreater–A–A–\textless blk\textgreater" are all semantically equivalent to the target label "A". This intrinsic equivalence provides a principled justification for pruning redundant non-blank frames without degrading decoding correctness.

\textbf{KOO's hidden gem}
To the best of our knowledge, existing End-to-End ASR technologies employing a TLG graph for decoding suffer from a latent issue during the construction of the static Token.fst graph \cite{ref26, ref27}: the probabilities of consecutive non-blank frames become excessively high due to self-loop operation at the non-initial states. To counteract the path probability inflation caused by consecutive identical non-blank frames, the KOO algorithm exclusively selects the single frame with the highest posterior probability during the FST decoding process. Tables \ref{tab:in-house} and \ref{tabel2} (G1 vs. G6) further validate the effectiveness of KOO in resloving the path probability inflation. These findings introduce a novel and effective paradigm for leveraging CTC/AED posterior probabilities during FST-based decoding, offering new insights for further optimization.

\section{Conclusion}
\label{sec:conclusion}

In this paper, we thoroughly explore the spiking behavior of CTC/AED outputs and propose the conjecture that blank frames provide positional information, while non-blank frames carry semantic information beneficial to the model. Building on this, we present two complementary algorithms to enhance both inference speed and recognition accuracy of CTC/AED-based E2E ASR systems named IOO and KOO algorithms. By reconstructing the sequence of blank and non-blank frames, our method enables a more efficient integration with WFSTs, drastically reducing the number of decoding frames. Additionally, the problem of severe accuracy degradation in AED-FST can be mitigated by using the IOO algorithm. 
Futhermore, we introduce a weight pushing optimization between the det and min steps, improving TLG search efficiency through early pruning. The experimental results on In-House, AISHELL-1 and Librispeech datasets confirm that the IOO and KOO algorithms can significantly enhance inference speed while even improving recognition accuracy.



\section*{Limitations}
While this study demonstrates significant improvements in decoding efficiency and accuracy, the evaluation is currently concentrated on standard benchmarks such as AISHELL-1 and LibriSpeech. Consequently, the robustness of the proposed algorithms across a broader spectrum of low-resource languages and complex acoustic environments remains to be fully explored.

\bibliography{latex/custom.bib}

\clearpage
\appendix
\section{IOO for CTC-FST}
\label{sec:Appendix}

\begin{algorithm}[H]
\caption{IOO and KOO Algorithms for CTC}
\label{alg:1}
\begin{algorithmic}[1] 
\footnotesize 
\setstretch{1.15} 

    \Require Posterior probability $P \in \mathbb{R}^{T \times V}$
    \Ensure Processed sequence $H_{final}$

    
    \Function{ProcessPosteriorsCTC}{$P$}
        \State $B \KwTo \mathrm{Queue}$ \Comment{Init frames block queue}
        \State $i,j,k \KwTo 0$
        \State $koo \KwTo $ false \Comment{Init koo setting}
        \While{$i < T$}{
            \State $p\_c \KwTo -1$ \Comment{Init previous $c$ with -1}
            \State $c, p \KwTo \argmax(P[i]), \max(P[i])$
            \If{$c \neq \mathrm{p\_c}$}
                \State $p\_c \KwTo c$
                \State $b \KwTo [(P[i], p)]$ \Comment{Init sub block $b$}
                \State $j \KwTo i + 1$
                
                \While{$j < T \land \argmax(P[j]) = c$}
                    \State $b.\text{append}((P[j], \max(P[j])))$
                    \State $j \KwTo j + 1$
                \EndWhile
                \State $B.\text{push}(c, b)$ \Comment{Push $b$ to queue $B$}
                \State $i \KwTo j$ \Comment{Jump to next block}
            \Else
                \State $i \KwTo i + 1$
            \EndIf
        }
        \EndWhile

        \State $H_{final} \KwTo []$
        \State $v_{blk} \KwTo [1.0, 0, \dots, 0]$ \Comment{One-hot blank vector}
        \State $H_{final}.\text{append}(v_{blk})$ \Comment{First customized frame}
        \vspace{0.05cm} 
        \While{$B$ is not empty}
            \State $b \KwTo B.\mathrm{pop}$
            \If{$k = \mathrm{0}  \land b.\mathrm{key} = \mathrm{0}$}
                \State $k \KwTo k + 1$ 
                \State $\mathrm{continue}$
            \EndIf
            \If{ $b.\mathrm{key} = $ 0}
                \State \textcolor{blue!70!black}{\textbf{Phase 1: Insert-Only-One (IOO)}} 
                \State $H_{final}.\text{append}(v_{blk})$ \Comment{With IOO}
            \ElsIf{$\mathrm{not} koo$}
                \State $H_{final}.\text{append}(b.\mathrm{value})$ \Comment{Without KOO}
            \Else
                \State \textcolor{blue!70!black}{\textbf{Phase 2: Keep-Only-One (KOO)}}
                \State $f^* \KwTo \operatorname{SelectMaxProbFrame}(b.\mathrm{value})$
                \State $H_{final}.\text{append}(f^*)$
            \EndIf
        \EndWhile    
        \State \Return $H_{final}$
    \EndFunction
\end{algorithmic}
\end{algorithm}

\section{IOO for AED-FST}
\begin{algorithm}[H]
\caption{IOO Algorithm for AED}
\label{alg:2}
\begin{algorithmic}[1] 
\footnotesize 
\setstretch{1.15} 

    \Require Posterior probability $P \in \mathbb{R}^{T \times V}$
    \Ensure Processed sequence $H_{final}$

    \vspace{0.1cm} 
    
    \Function{ProcessPosteriorsAED}{$P$}
        \State $i \KwTo 0$
        \State $H_{final} \KwTo []$
        \State $v_{blk} \KwTo [1.0, 0, \dots, 0]$ \Comment{One-hot blank vector}
        \State $H_{final}.\text{append}(v_{blk})$
        \While{$i < T$}
            \State $H_{final}.\text{append}(P[i])$
            \State $H_{final}.\text{append}(v_{blk})$
            \State $i \KwTo i + 1$
        \EndWhile
        \State \Return $H_{final}$
    \EndFunction

\end{algorithmic}
\end{algorithm}

\section{Encoder Configurations}
\begin{table}[h]
\caption{Zipformer configurations used in our experiments.}
\centering
\setlength{\tabcolsep}{0.1mm}{
\small
\begin{tabular}{ll}
\toprule
\textbf{Parameter} & \textbf{Values} \\
\midrule
Layers num & 2, 2, 4, 6, 4, 2 \\
FFN dim   & 512, 768, 1536, 2048, 1536, 768 \\
Encoder dim      & 192, 256, 512, 768, 512, 256 \\
Encoder-unmasked dim & 192, 192, 256, 320, 256, 192 \\
\bottomrule
\end{tabular}}
\label{encoder_config}
\end{table}

\section{KOO for Non-Blank Frames}
\begin{table}[h]
    \caption{Results of thresholds of non-blank frames on the In-House dataset.}
    \renewcommand{\arraystretch}{1.0}
    \centering
    \small
    \label{non-blank}
    \setlength{\tabcolsep}{1.5mm}{
    \begin{tabular}{c c c c}
    \toprule
    \textbf{ID} & \textbf{Non-balnk threshold}  & \textbf{CER (\%)} $\downarrow$ & \textbf{Speed up} $\uparrow$ \\
    \midrule
    \textbf{G3} & $\mathrm{\{* \}}$ & 3.78 &  1.72 $\times$ \\
    \textbf{G7}  & $\mathrm{\{*\ge0.8\}}$ & 3.77 &  1.73 $\times$ \\
    \textbf{G8} & $\mathrm{\{*\ge0.90\}}$ & 3.29 &  1.62 $\times$ \\
    \textbf{G9}  & $\mathrm{\{*\ge0.95\}}$ & 3.13 &  1.64 $\times$ \\
    \textbf{G10} & $\mathrm{\{*\ge0.99\}}$ & 3.06 &  1.61 $\times$ \\
    \bottomrule
    \end{tabular}}
\end{table}


\end{document}